\documentclass[seceq]{ptptex}
\usepackage[dvips]{graphics}
\usepackage{graphicx}
\usepackage{amssymb}
\usepackage{amsmath}
\usepackage{array}
\usepackage{bm}
\usepackage[logonly]{trace}

\title{ 
International Comparison of Labor Productivity Distribution for Manufacturing and Non-Manufacturing Firms
}

\author{ 
Yuichi \textsc{Ikeda}$^1$ and Wataru \textsc{Souma}$^2$
}

\inst{
$^1$ Hitachi Research Institute and $^2$ NiCT/ATR
}

\abst{
Labor productivity was studied at the microscopic level in terms of distributions based on individual firm financial data from Japan and the US. A power-law distribution in terms of firms and sector productivity was found in both countries' data. The labor productivities were not equal for nation and sectors, in contrast to the prevailing view in the field of economics. It was found that the low productivity of the Japanese non-manufacturing sector reported in macro-economic studies was due to the low productivity of small firms.
}

\begin{document}
\maketitle

\section{Introduction} 
\label{section:Introduction}
The Japanese non-manufacturing sector, which is the service sector in the extended sense of the meaning, was reported to have low productivity in economic studies \cite{Fukao2004, Mothashi2007}. Many researchers have cited low productivity as a serious problem of the Japanese economy. For this reason, policy makers, business administrators, and physicists have taken considerable interest in the reasons behind these figures.

A recent macroeconomic study \cite{JPCFSED2007} ranked the labor productivity of all industries in FY2005 (Table $\ref{tab:Table1}$). Japan (JP) ranked 20th, or about $71\%$ that of the United States (US). Moreover, Japanese manufacturers' productivity ranked 6th, or about $89\%$ of the US ($\ref{tab:Table2}$). The labor productivity of Japanese non-manufacturers is estimated to be about $61\%$ of US, from the fact that $64\%$ of Japanese GDP is created by non-manufacturers (see Table $\ref{tab:Table1}$ and $\ref{tab:Table2}$ ($0.36\times89\%+0.64x=71\%$ leads to $x=61\%$)).

\begin{table} 
\begin{center} 
\caption{Labor Productivity of All Industries in FY2005} 
\label{tab:Table1} 
\begin{tabular}{ccc} 
\hline \hline 
Ranking & Country & Productivity (\$/Worker) \\ 
\hline 
1st & Luxembourg & $104,610$ \\
2nd & Norway & $97,275$ \\ 
3rd & US & $86,714$ \\ 
$\dots$ & $\dots$ & $\dots$ \\ 
20th & Japan & $61,862$ ($71\%$ of US) \\ 
\hline \hline 
\end{tabular} 
\end{center} 
\end{table}

\begin{table} 
\begin{center} 
\caption{Labor Productivity of Manufacturers in FY2005} 
\label{tab:Table2} 
\begin{tabular}{ccc} 
\hline \hline Ranking & Country & Productivity (\$/Worker) \\ 
\hline 1st & Ireland & $170,872$ \\ 2nd & Norway & $97,733$ \\ 
3rd & US & $96,962$ \\ 
$\dots$ & $\dots$ & $\dots$ \\ 
6th & Japan & $86,608$ ($89\%$ of US) \\ 
\hline \hline 
\end{tabular} 
\end{center} 
\end{table}

In light of this background, we decided to study the labor productivity distribution at the microscopic level in terms of distributions based on individual firm financial data for Japan and the US. Japan's labor productivity and US labor productivity were compared for the manufacturing and non-manufacturing sectors.

In the following, first, the economic theory of productivity is briefly described. Then, results of data analyses of the labor productivity distribution are presented. Finally, the discrepancy between the results of the data analysis and economic theory is discussed.

\section{Economic Theory of Productivity}
The theory of productivity is briefly described to clarify what is expected for the labor productivity in the field of economics \cite{Stiglitz2002, Stiglitz2005}. We consider a firm's labor productivity to explain the theory. The following explanation is also valid for the industrial sector. Operating profit $\Pi$ is defined by 

\begin{equation} 
\Pi = p Y - r K - w L. 
\label{eq:TheoryProd1} 
\end{equation}
Here, $p$, $Y$, $r$, $K$, $w$, and $L$ are price, added value, interest rate, capital, wage rate, and labor, respectively. Each firm maximizes its profit $\Pi$ by adjusting $L$: 

\begin{equation} 
\frac{\partial \Pi}{\partial L} = p \frac{\partial Y}{\partial L} - w = 0. 
\label{eq:TheoryProd2} 
\end{equation}
Thus, marginal labor productivity $\partial Y / \partial L$ satisfies the relation:

\begin{equation} 
\frac{\partial Y}{\partial L} = \frac{w}{p}. 
\label{eq:TheoryProd3} 
\end{equation}
In equilibrium, the actual wage rate $w/p$ is equal for each firm, which means that there is no arbitrage opportunity for wage rates. Labor moves from firms offering low wages to firms offering high wages by an amount $\Delta L$, as shown in Figure \ref{fig:TheoryProd}, where the origin of labor of the $i^{th} (j^{th})$ firm $O_i (O_j)$ is the left (right) axis. As a result, the following relation is obtained for firm $i$ and firm $j$:

\begin{figure} 
\begin{center} 
\includegraphics[width=0.8\textwidth,bb=0 0 500 250]{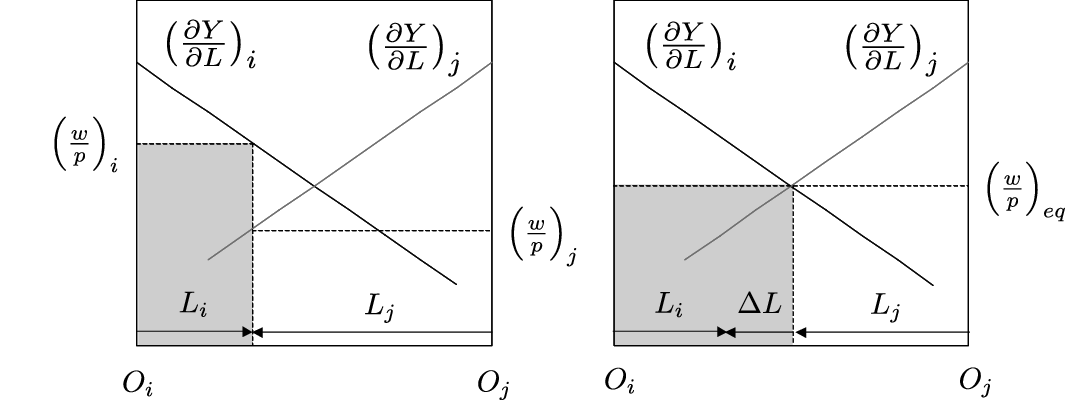}
\caption{In equilibrium (the right panel), the actual wage rate $w/p$ is equal for each firm, which means that there is no arbitrage opportunity for wage rates . Labor moves from firms offering low wages to firms offering high wages by an amount $\Delta L$} 
\label{fig:TheoryProd} 
\end{center} 
\end{figure}

\begin{equation} 
\left( \frac{\partial Y}{\partial L} \right)_{i} = \left( \frac{\partial Y}{\partial L} \right)_{j}. 
\label{eq:TheoryProd4} 
\end{equation}
If added value $Y$ is described using the Cobb-Dauglass production function \cite{CD1928},

\begin{equation} 
Y = A K^\alpha L^\beta, 
\label{eq:TheoryProd5} 
\end{equation}
where $A$, $\alpha$, $\beta$ are fitting parameters, the marginal labor productivity $\partial Y / \partial L$ can be rewritten using labor productivity $Y/L$:

\begin{equation} 
\frac{\partial Y}{\partial L} = \beta \frac{Y}{L}. 
\label{eq:TheoryProd6} 
\end{equation} 
Thus, labor productivity satisfies the following relation for firms $i$ and $j$:
\begin{equation} 
\beta_i \left( \frac{Y}{L} \right)_{i} = \beta_j \left( \frac{Y}{L} \right)_{j}. 
\label{eq:TheoryProd7} 
\end{equation}
There is a drawback to the use of a simple production function (\ref{eq:TheoryProd5}). If constant returns to scale $\alpha + \beta = 1$ is assumed, the following relation between productivity $Y/L$ and capital equipment ratio $K/L$ is obtained:

\begin{equation} 
\frac{Y}{L} = A \left( \frac{K}{L} \right)^\alpha. 
\label{eq:ProdCapEquip} 
\end{equation}
This means that a simple production function provides no information on the size dependence of the productivity.

\section{Data Analysis of Productivity Distribution}
Labor productivity distributions and Pareto indexes were analyzed for JP and US listed firms. Productivities of manufacturing and non-manufacturing firms were compared. Analyzed financial data were Bloomberg data from 1990 to 2003. The analysis used two different definitions of labor productivity:

\begin{equation} 
Labor~Productivity = \frac{Gross~Margin}{Worker}, 
\label{eq:LaborProd1} 
\end{equation}

\begin{equation} 
Labor~Productivity = \frac{Added~Value}{Worker}. 
\label{eq:LaborProd2} 
\end{equation}
Here, gross margin and added value are defined by

\begin{equation} 
Gross~Margin = Revenue - Cost~of~Goods~Sold, 
\label{eq:LaborProd3} 
\end{equation}

\begin{equation} 
Added~Value = Gross~Margin + Total~Labor~Cost = \frac{Gross~Margin}{1 - Labor~Share}, 
\label{eq:LaborProd4} 
\end{equation}
respectively. Note that labor share is a macro-economic quantity. Workers do not include part-time and contract workers.

\begin{figure} 
\begin{center} 
\includegraphics[width=0.45\textwidth,bb=0 0 500 440]{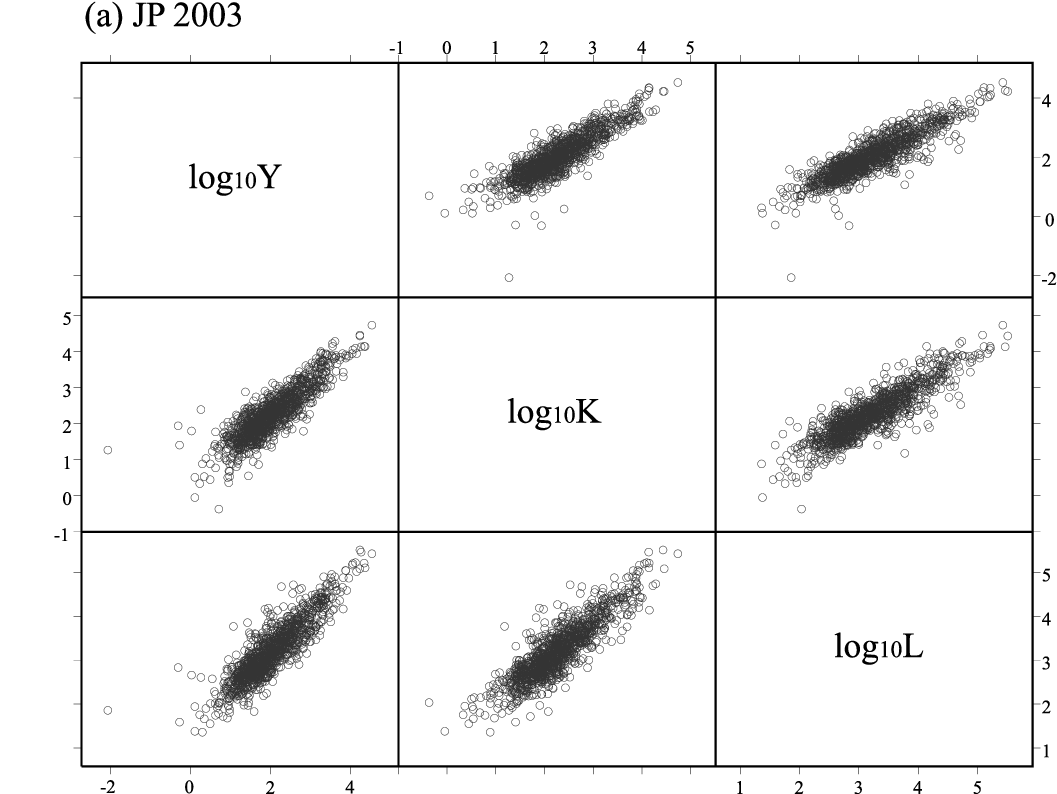}
\includegraphics[width=0.45\textwidth,bb=0 0 500 440]{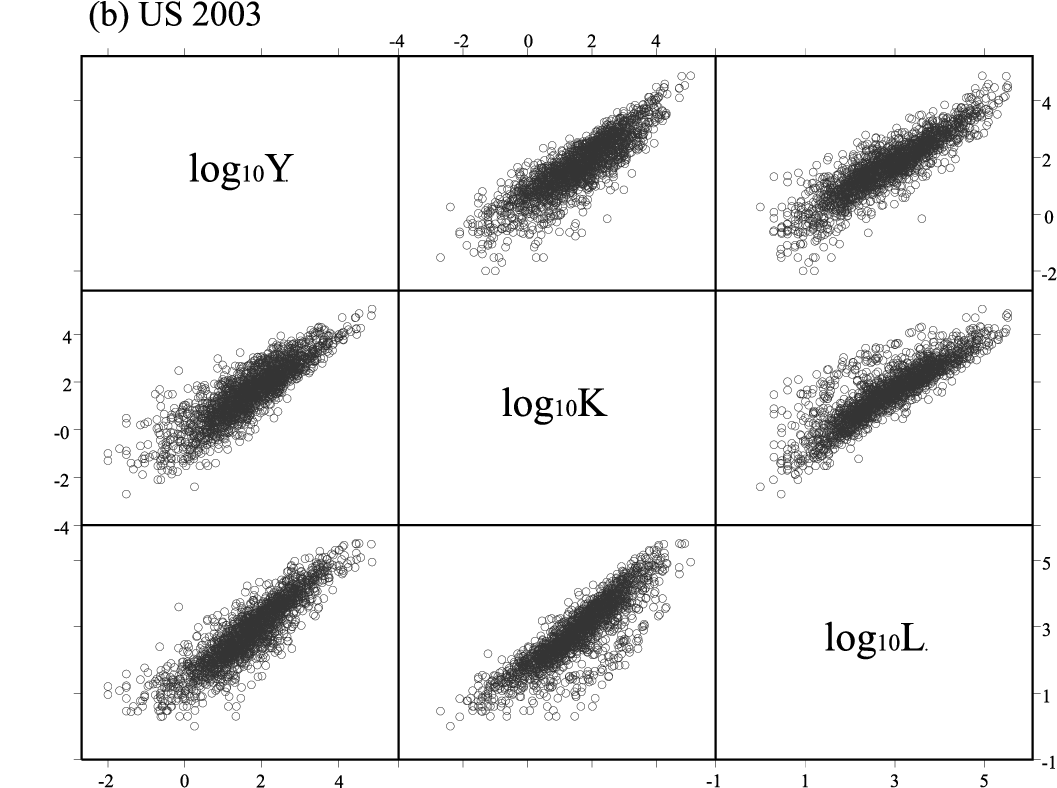}
\caption{Scatter plots of gross margin $Y$, capital $K$, and labor $L$ show linear correlations for log transformed variables of gross margin $Y$, capital $K$, and labor $L$.} 
\label{fig:YKL2003M} 
\end{center} 
\end{figure}

The parameters of the microscopic production function were estimated for JP and US listed firms using the Cobb-Douglass production function (\ref{eq:TheoryProd5}). By taking log transformation of variables in Eq. (\ref{eq:TheoryProd5}), we obtain the relation,

\begin{equation} 
\log_{10} Y_t = \log_{10} A + \alpha \log_{10} K_t + \beta \log_{10} L_t. 
\label{eq:LogCobbDouglas} 
\end{equation}
Scatter plots of gross margin $Y$, capital $K$, and labor $L$ are shown in Figure \ref{fig:YKL2003M}. Linear correlations are observed for log transformed variables of gross margin $Y$, capital $K$, and labor $L$. This means that the Cobb-Douglass production function is an appropriate functional form. Parameters $A$, $\alpha$, and $\beta$ were estimated using multi-regression analysis.

\begin{figure} 
\begin{center} 
\includegraphics[width=0.45\textwidth,bb=0 0 500 440]{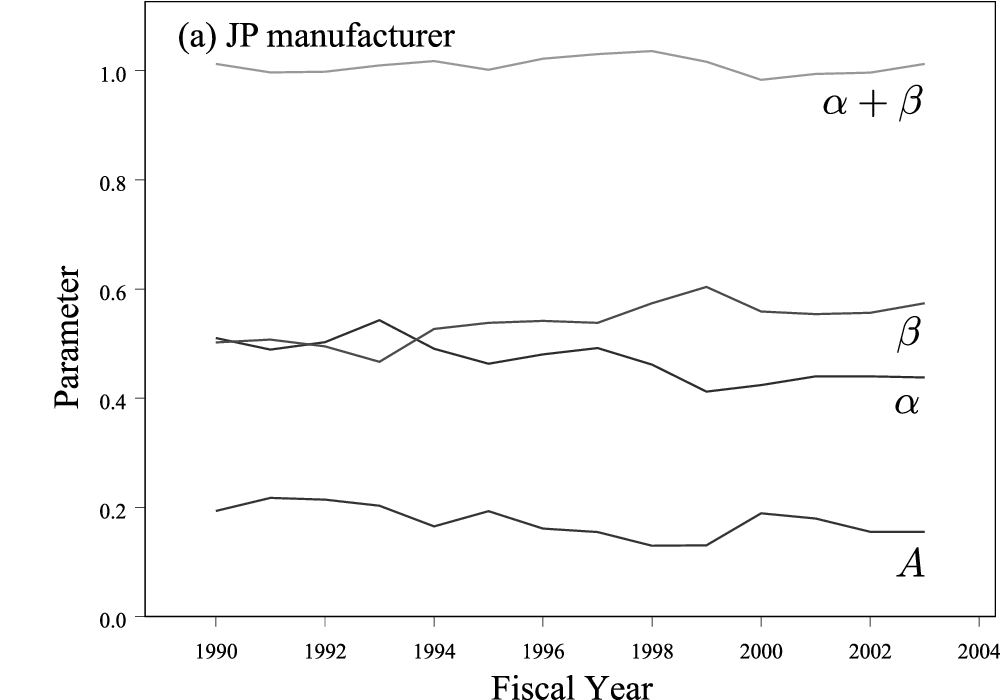} 
\includegraphics[width=0.45\textwidth,bb=0 0 500 440]{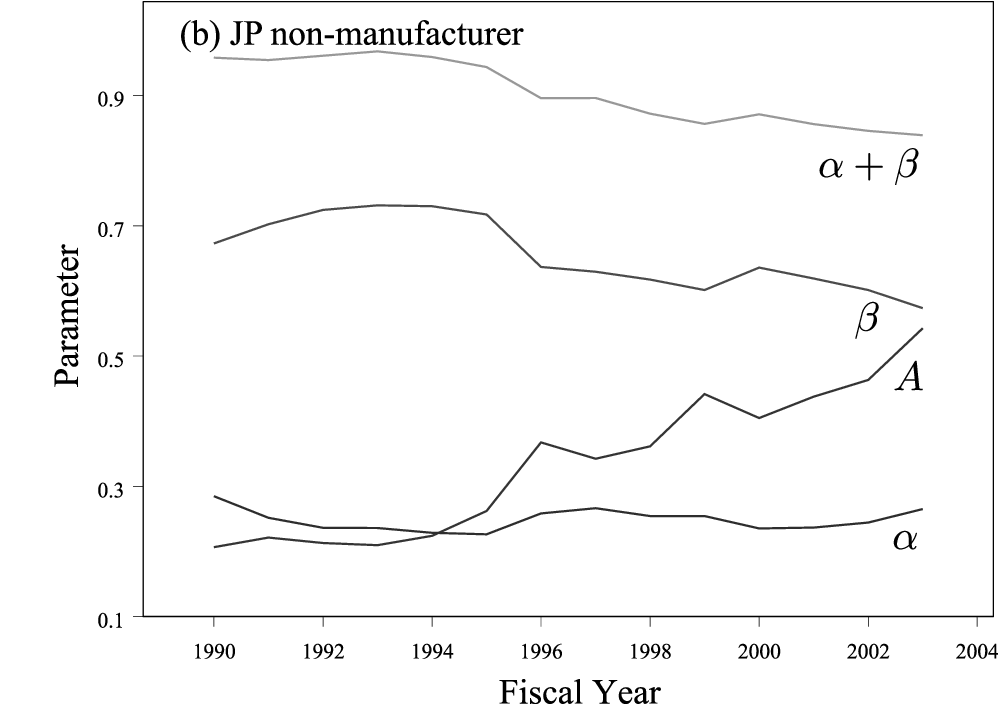} 
\includegraphics[width=0.45\textwidth,bb=0 0 500 440]{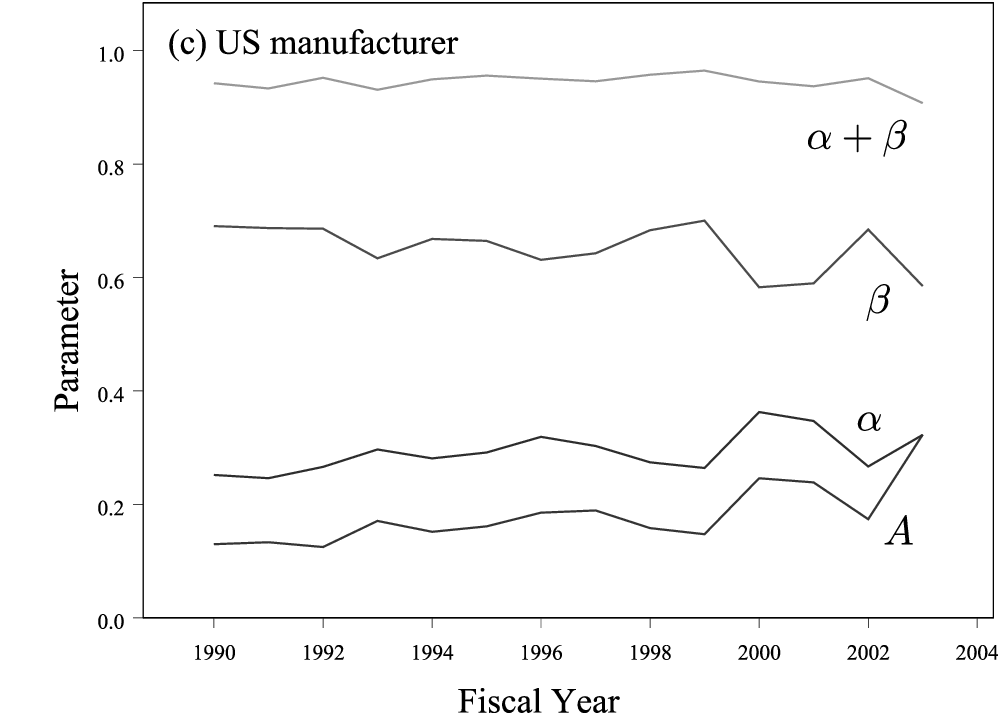} 
\includegraphics[width=0.45\textwidth,bb=0 0 500 440]{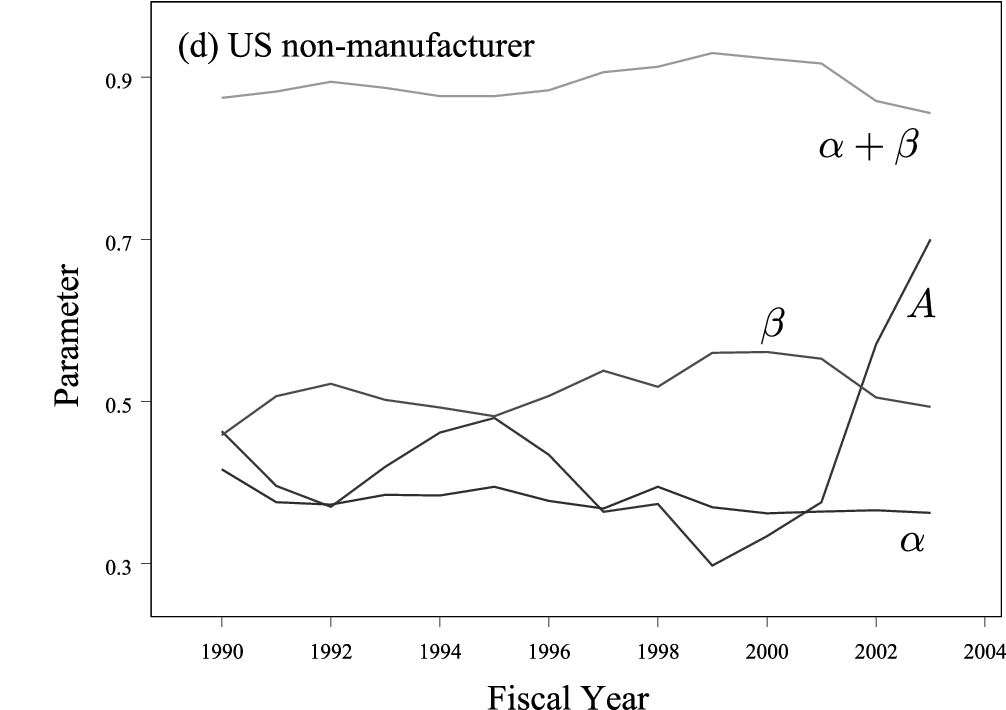} 
\caption{Parameters $A$, $\alpha$, and $\beta$ were estimated using multi-regression analysis with the Cobb-Douglass production function. The obtained parameters of the production function are shown. JP and US manufacturers have constant returns to scale ($\alpha + \beta = 1$). On the other hand, JP and US non-manufacturers have decreasing returns to scale ($\alpha + \beta < 1$).} 
\label{fig:ProdFunc} 
\end{center} 
\end{figure}

Figure \ref{fig:ProdFunc} shows the parameters obtained for the production function. Manufacturers (both JP and US) have constant returns to scale ($\alpha + \beta = 1$). On the other hand, non-manufacturers (both JP and US) have decreasing returns to scale ($\alpha + \beta < 1$). Note that $\beta = 0.6$ for all cases. Thus, from Eq. (\ref{eq:TheoryProd7}), we can say that labor productivities $Y/L$ for firm $i$ and firm $j$ are equal if an equilibrium state is achieved.

\begin{figure} 
\begin{center} 
\includegraphics[width=0.45\textwidth,bb=0 0 500 440]{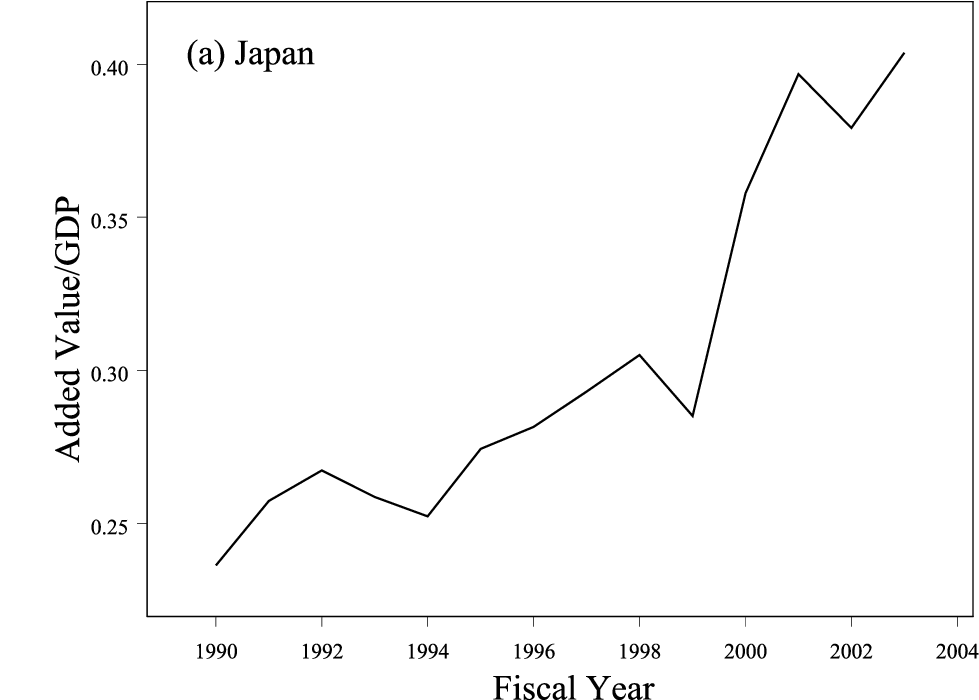} 
\includegraphics[width=0.45\textwidth,bb=0 0 500 440]{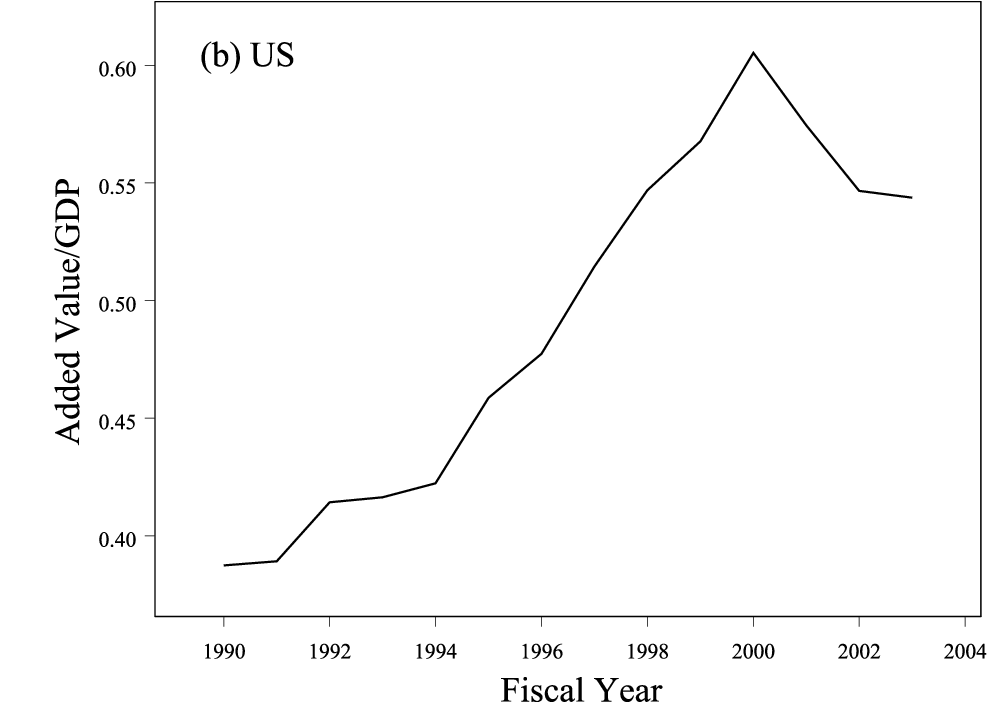} 
\caption{Aggregated added value created by listed firms in JP and the US. The vertical axis is aggregated added value divided by gross domestic product (GDP). About half of the GDP is created by listed firms in the JP and US.} 
\label{fig:AggregatedAddedValue} 
\end{center} 
\end{figure}

Next, the labor productivity of firms was analyzed. Figure \ref{fig:AggregatedAddedValue} plots the aggregated added value created by the listed firms. The vertical axis is aggregated added value divided by gross domestic product (GDP). About half of the GDP is created by listed firms in JP and the US. This means that a large fraction of economy of each nation is covered by the analysis of the listed firms.

\begin{figure} 
\begin{center} 
\includegraphics[width=0.45\textwidth,bb=0 0 500 440]{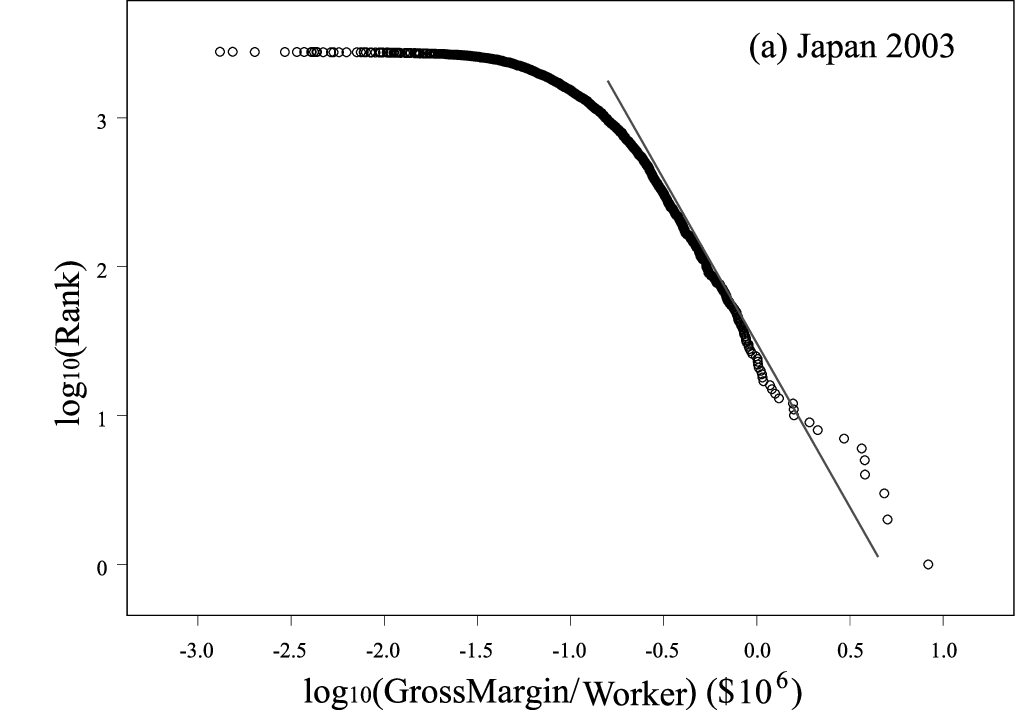} 
\includegraphics[width=0.45\textwidth,bb=0 0 500 440]{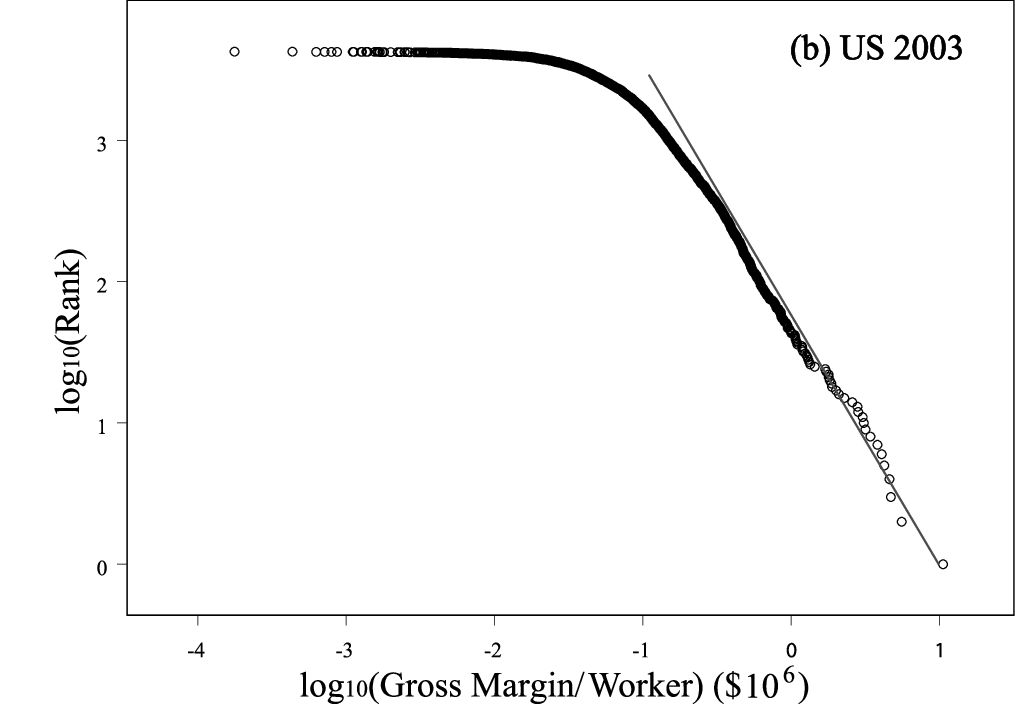} 
\caption{Rank-size plot of labor productivity for firms. Because the horizontal and vertical axes are log scales, the straight section indicates a power-law distribution.} 
\label{fig:ProdFirms} 
\end{center} 
\end{figure}

Figure \ref{fig:ProdFirms} represents the rank-size plot of labor productivity for firms. Because the horizontal and vertical axes are log scales, the straight section indicates a power-law distribution. The tail of the labor productivity distribution for firms is the power-law distribution, 
\begin{equation} 
P_>(x) \propto x^{- \mu}, 
\label{eq:Power-LawDist} 
\end{equation}
where $\mu$ is called the Pareto index. The Pareto index $\mu$ was estimated using a least squares fitting to the tail of the productivity distribution.

\begin{figure} 
\begin{center} 
\includegraphics[width=0.45\textwidth,bb=0 0 500 440]{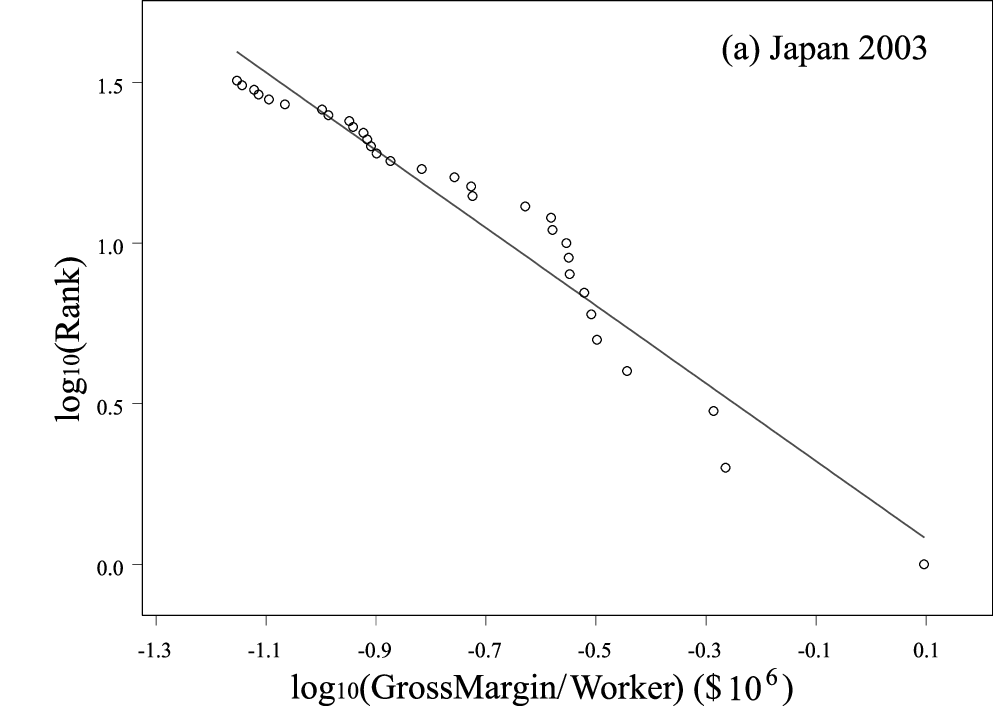} 
\includegraphics[width=0.45\textwidth,bb=0 0 500 440]{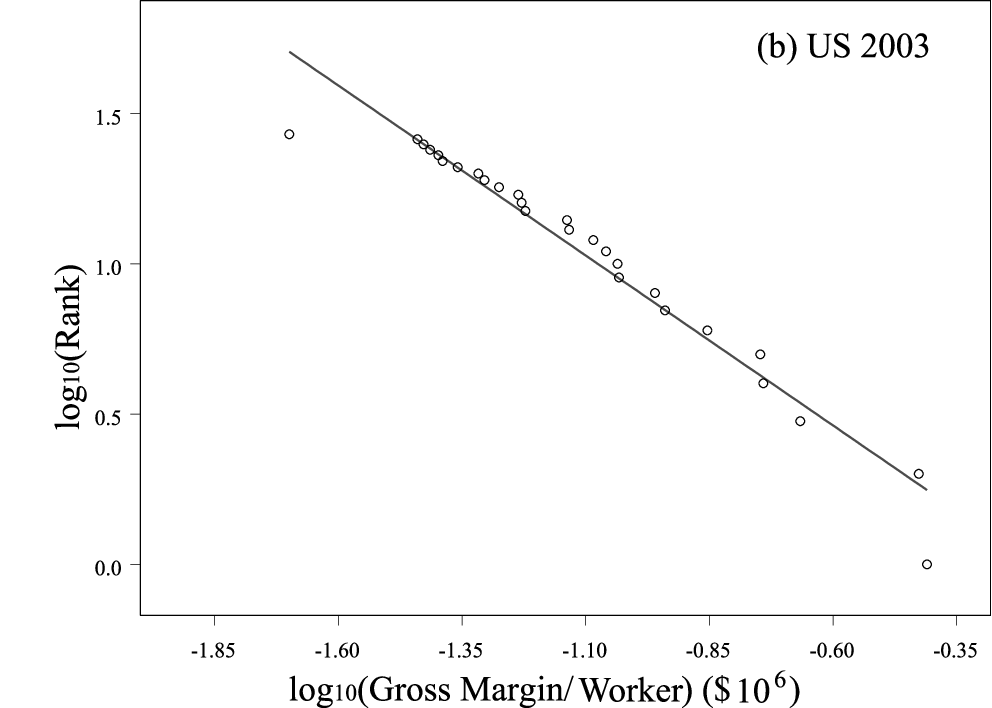} 
\caption{Rank-size plot of labor productivity for industrial sectors. Each circle indicates an individual sector. Labor productivity for industrial sectors follows a power-law distribution.} 
\label{fig:ProdSectors} 
\end{center} 
\end{figure}

The labor productivity of the industrial sector was analyzed. Manufacturers included food, textile, pulp and paper, chemical, pharmaceutical, petroleum, steel, nonferrous metals, machinery, electrical machinery, automobile, shipbuilding, transportation equipment, precision apparatus, rubber, and ceramics. Non-manufacturers included fisheries, mining, construction, telecommunications, road transportation, railroads, marine transportation, air transportation, warehousing, wholesale, retail, service, electric power, gas, and real estate. The rank-size plots of labor productivity for industrial sectors are shown in Figure \ref{fig:ProdSectors}. Each circle indicates an individual sector. Labor productivity for industrial sectors follows a power-law distribution. The Pareto index was estimated using the least squares method on the whole distribution.

The observed power-law distribution for firms and sectors means that an equilibrium state is not achieved, in contrast to what is expected in economics, because the power law distribution is, in general, obtained for the critical state, not for the equilibrium state. This discrepancy should be recognized as a serious fault of neoclassical economics. 

\begin{figure} 
\begin{center} 
\includegraphics[width=0.45\textwidth,bb=0 0 500 440]{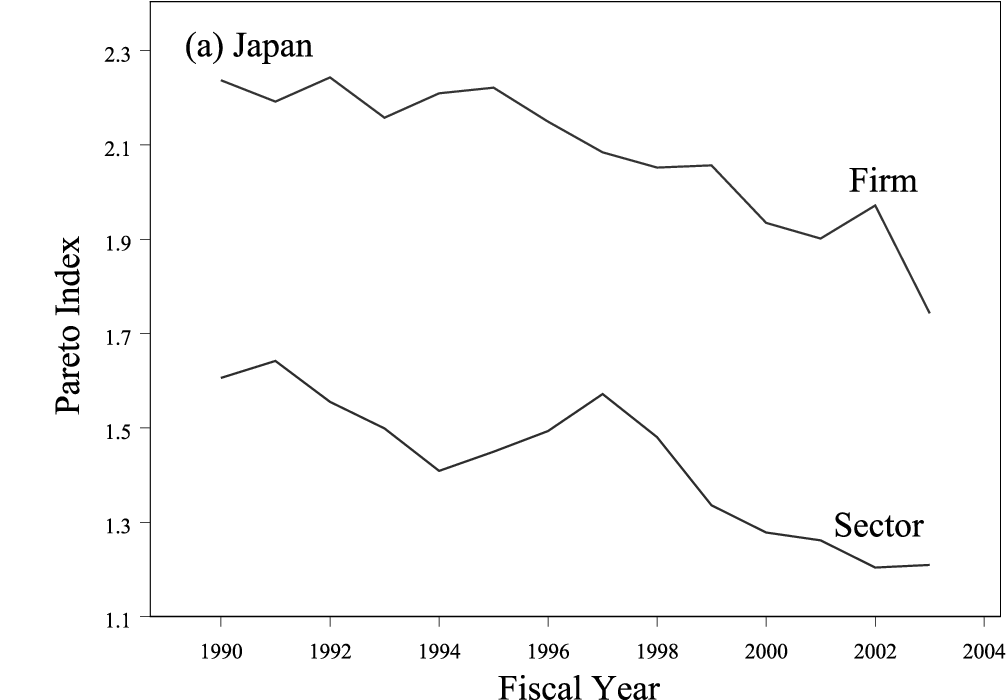} 
\includegraphics[width=0.45\textwidth,bb=0 0 500 440]{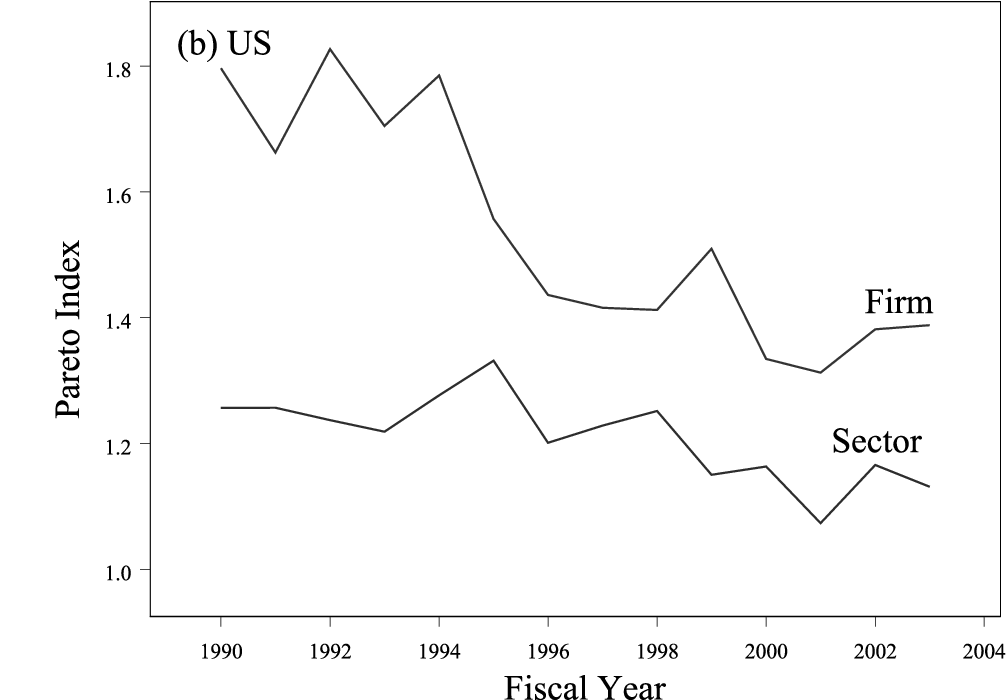} 
\caption{Time variation of the estimated Pareto indices for firms and sectors from 1990 to 2003. Pareto indices for firms are always larger than the indices for the industrial sector.} 
\label{fig:ParetoIndices} 
\end{center} 
\end{figure}

Figure \ref{fig:ParetoIndices} shows the time variation of the estimated Pareto indices for firms and sectors from 1990 to 2003. Pareto indices for firms are always larger than the indices for industry sectors. Pareto indices for the US is smaller than those for JP. In particular, the index is for US firms is small after 1994, which is the year the World Trade Organization was established and globalization of the economy began. This suggests that the economic disparity in the US grew as a result of globalization.

\begin{figure} 
\begin{center} 
\includegraphics[width=0.45\textwidth,bb=0 0 500 440]{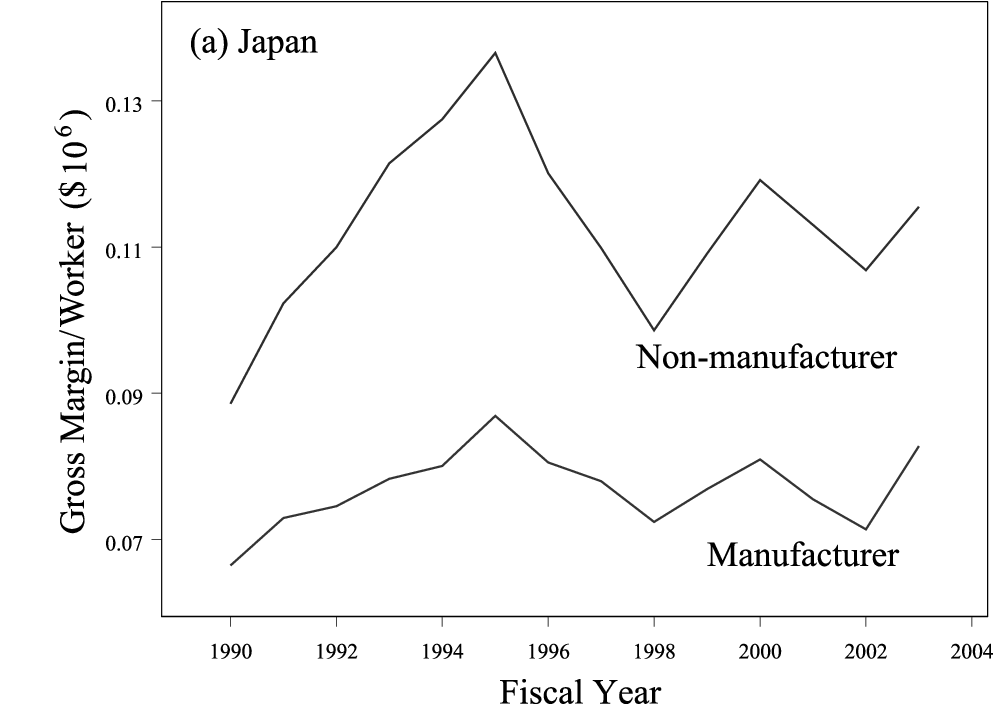} 
\includegraphics[width=0.45\textwidth,bb=0 0 500 440]{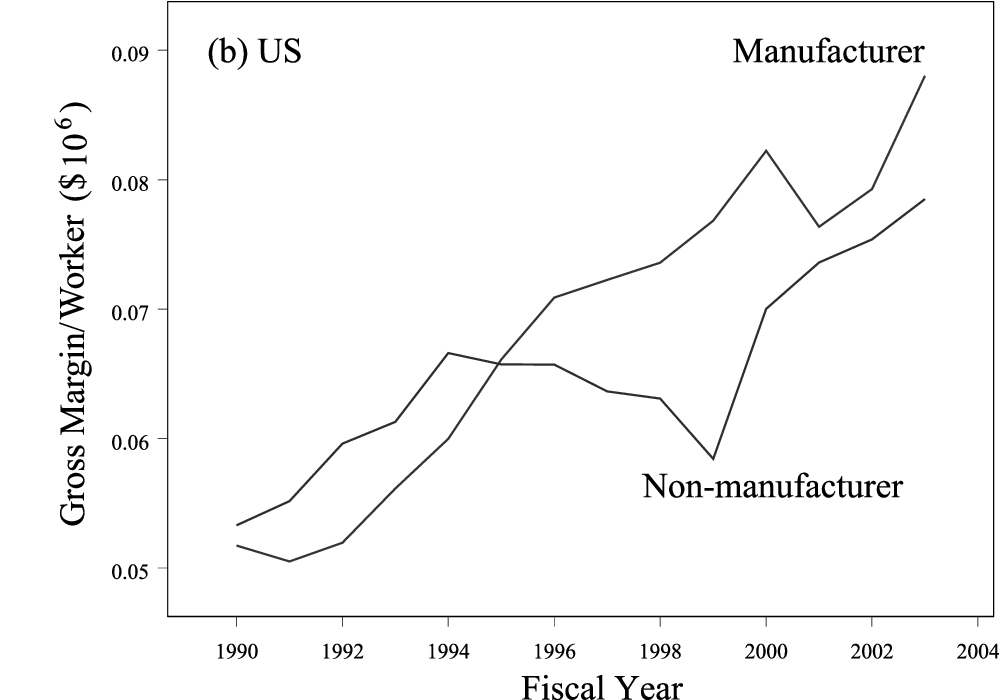} 
\caption{Time variation of labor productivity of manufacturers and non-manufacturers from 1990 to 2003. It is expected from Eq. (\ref{eq:TheoryProd7}) with $\beta = 0.6$ that labor productivities $Y/L$ are equal for firms and sectors. The labor productivity of JP non-manufacturers is higher than that of US non-manufacturers in the range of listed firms.} 
\label{fig:LaborProd} 
\end{center} 
\end{figure}

Figure \ref{fig:LaborProd} shows the time variation of labor productivity of manufacturers and non-manufacturers from 1990 to 2003. It is expected from Eq. (\ref{eq:TheoryProd7}) with $\beta = 0.6$ that labor productivities $Y/L$ should be equal for firms and sectors. However, the obtained results show differences between nation and sectors. Specifically, the labor productivity of JP non-manufacturers is higher than that of US non-manufacturers in the range of listed firms. This result seems to contradict the results of the macroeconomic studies mentioned in section \ref{section:Introduction}.

By assuming that both our result and the macro-economic studies are correct, we can hypothesize from the study of listed firms that the low productivity of JP non-manufacturers reported in the macro-economic studies has its root in the low productivity of unlisted small firms. To test this hypothesis, we analyzed financial data consisting of listed and unlisted JP firms. Analyzed data were combined data of Nikkei NEEDS and credit risk database (CRD) \cite{CRD}. Nikkei NEEDS includes only listed firms, and CRD includes only unlisted small firms. The combined data includes $4 \times 10^5$ firms approximately. Although this analysis used a different definition of added value: 

\begin{multline} 
Added~Value = Ordinary~Income + Total~Labor~Cost + \\ Financial~Expense + Tax~and~Public~Charge + Depreciation~Cost, 
\label{eq:AddedValue} 
\end{multline}
it gives the same result as the the previous definition gives.

\begin{figure} 
\begin{center} 
\includegraphics[width=0.45\textwidth,bb=0 0 500 460]{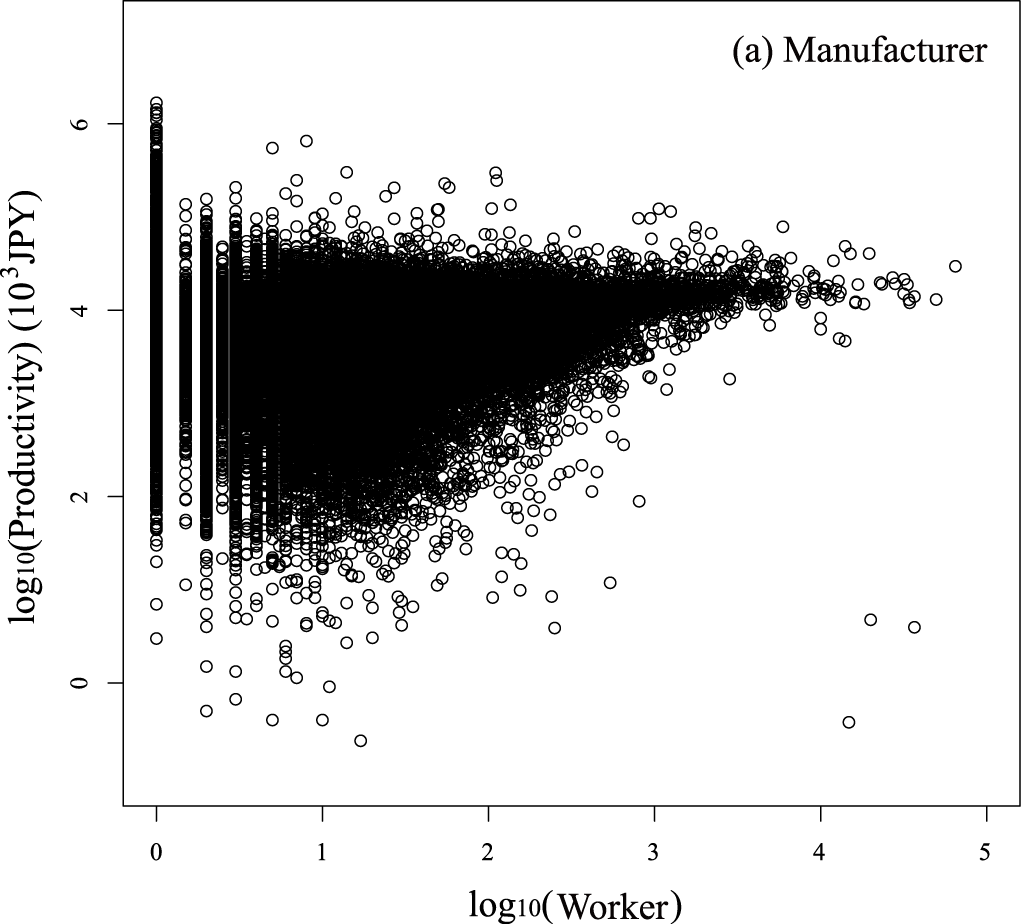} 
\includegraphics[width=0.45\textwidth,bb=0 0 500 460]{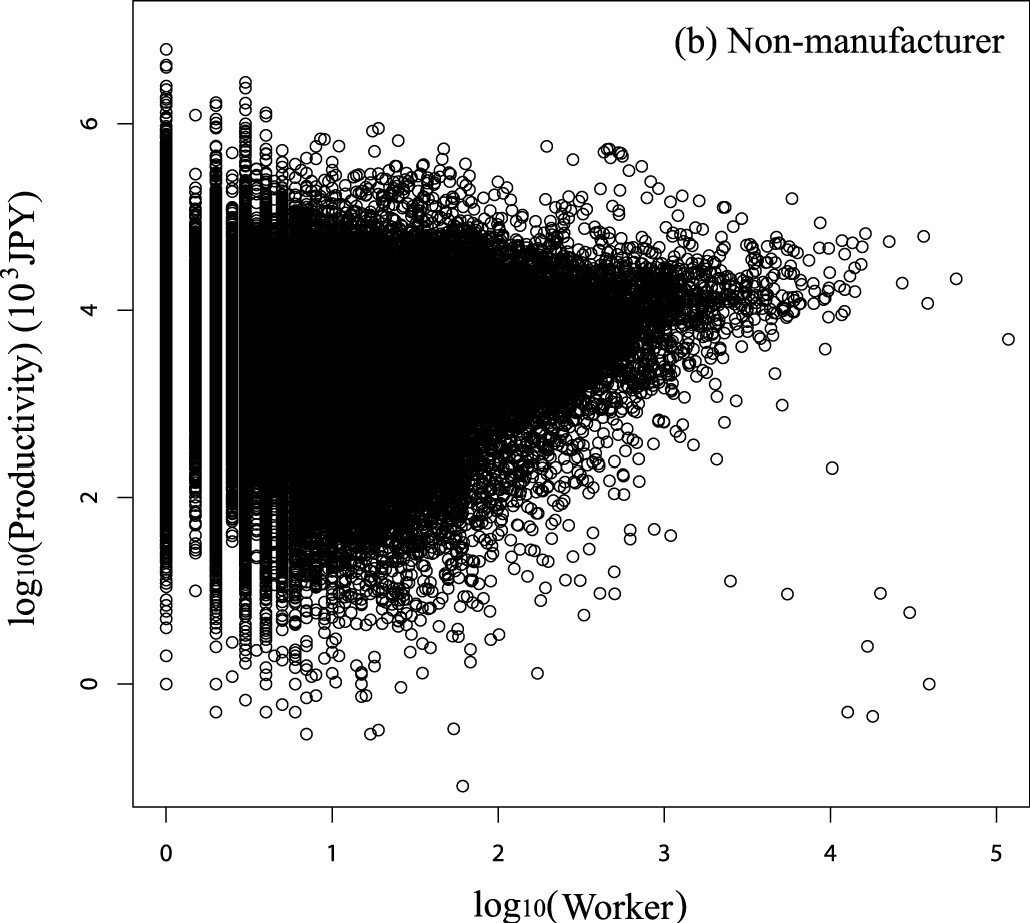} 
\caption{Productivities vs number of workers in 2005 for manufacturers and non-manufacturers. The productivity of large non-manufacturers is at the same level as that of large manufacturers. But the productivity of small non-manufacturers is more widely distributed than that of small manufacturers.} 
\label{fig:ProdVsEmpl} 
\end{center} 
\end{figure}

Figure \ref{fig:ProdVsEmpl} plots productivity vs number of workers in 2005. The productivity of large non-manufacturers is at the same level as that of large manufacturers. But the productivity of small non-manufacturers is more widely distributed than that of small manufacturers. These figures indicate that there is a size dependence to productivity, which is not accounted for in Eq. (\ref{eq:ProdCapEquip}).

\begin{figure} 
\begin{center} 
\includegraphics[width=0.45\textwidth,bb=0 0 500 440]{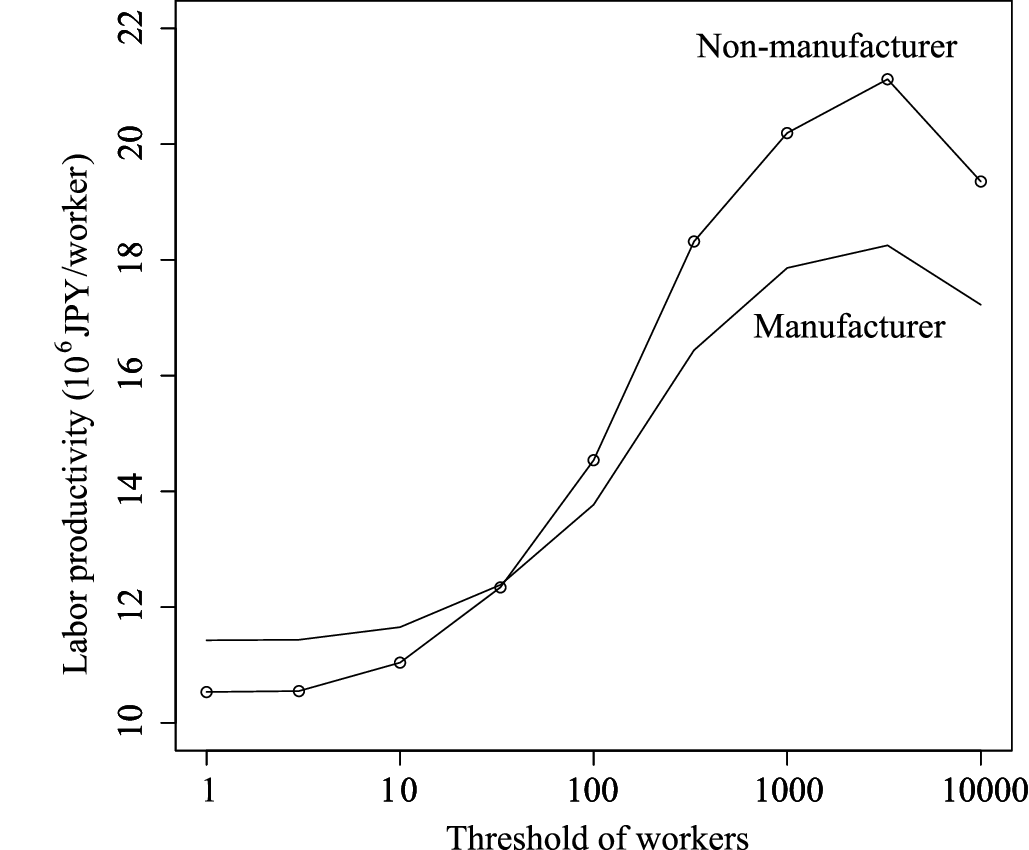} 
\caption{Size dependence of productivity. Lower productivity for smaller firms is apparent. The low productivity of the Japanese non-manufacturing sector is due to the low productivity of small firms.} 
\label{fig:Size-dependence} 
\end{center} 
\end{figure}

Figure \ref{fig:Size-dependence} shows the size dependence of productivity in a different way. The vertical axis is productivity obtained for firms with work forces larger than a threshold. Lower productivity for smaller firms is apparent. The low productivity of the Japanese non-manufacturing sector is thus due to the low productivity of small firms. This is the origin of the contradiction pointed out above.

The above results reveal the need for an elaborate modeling method, rather than a simple production function, to analyze the size dependence of productivity. The copula method \cite{Nelsen2006} is suitable to model the whole distribution of ($Y$, $K$, $L$) by taking into account nonlinear correlations.

\section{Conclusion}
Labor productivity was studied at the microscopic level in terms of distributions based on individual firm financial data for Japan and the US.

A power-law distribution for firms and sector productivity was observed in both countries' data. The power-law distribution for firms and sectors means that an equilibrium state is not achieved. The labor productivities were not equal for nation and sectors, in contrast to what can be expected from Eq. (\ref{eq:TheoryProd7}) with $\beta = 0.6$. It should be noted that the observed non-equlibrium state is evidence against the prevailing view in economics. 

The data analysis shows that the labor productivity of JP non-manufacturers is higher than that of US non-manufacturers in the range of listed firms. It was clarified that the low productivity of the Japanese non-manufacturing sector reported in macro-economic studies was caused by the low productivity of small firms.

\section*{Acknowledgements}
The authors would like to thank Prof.~H.~Aoyama (Kyoto University), Prof.~H.~Iyetomi (Niigata University) and Dr.~Y.~Fujiwara (NiCT/ATR) for their enlightening discussions. The authors also thank the Yukawa Institute for Theoretical Physics at Kyoto University. Discussions held during the YITP workshop YITP-W-07-16 on "Econophysics III -Physical Approach to Social and Economic Phenomena-" were invaluable to completing this work.

\end{document}